\documentclass[letterpaper, 10 pt, conference]{ieeeconf}
\usepackage[utf8]{inputenc}
\usepackage{cite}
\usepackage{amsmath,amssymb,amsfonts}
\usepackage{graphicx}
\usepackage{subfig}
\usepackage{textcomp}
\usepackage{xcolor}
\usepackage{algorithm}
\usepackage{algorithmicx}
\usepackage{algpseudocode}
\usepackage{bm}
\usepackage{booktabs}
\usepackage{wrapfig}
\usepackage{tabularray}
\usepackage{url}
\newtheorem{theorem}{Theorem}
\newtheorem{prop}[theorem]{Lemma}
\usepackage{array}
\newcolumntype{x}[1]{>{\centering\arraybackslash\hspace{0pt}}p{#1}}

\usepackage{times}

\IEEEoverridecommandlockouts
\begin{document}

\title{\LARGE \bf
 Pontryagin Optimal Control via Neural Networks
}

\author{Chengyang Gu,~\IEEEmembership{Student Member, IEEE}, Hui Xiong,~\IEEEmembership{Fellow, IEEE} and Yize Chen,~\IEEEmembership{Member, IEEE}$^{1}$
\thanks{The authors are with the Artificial Intelligence Thrust, Hong Kong University of Science and Technology (Guangzhou), China. Emails: {\tt\small cgu893@connect.hkust-gz.edu.cn, \{xionghui, yizechen\}@ust.hk }.}%
}%

\maketitle

\begin{abstract}%
Solving real-world optimal control problems are challenging tasks, as the complex, high-dimensional  system dynamics are usually unrevealed to the decision maker. It is thus hard to find the optimal control actions numerically. To deal with such modeling and computation challenges, in this paper, we integrate Neural Networks with the Pontryagin's Maximum Principle (PMP), and propose a sample efficient framework Neural-PMP. The resulting controller can be implemented for systems with unknown and complex dynamics. By taking an iterative approach, the proposed framework not only utilizes the accurate surrogate models parameterized by neural networks, it also efficiently recovers the optimality conditions along with the optimal action sequences via PMP conditions. Numerical simulations on Linear Quadratic Regulator, energy arbitrage of grid-connected lossy battery, control of single pendulum, and two MuJoCo locomotion tasks demonstrate our proposed Neural-PMP is a general and versatile computation tool for finding optimal solutions. And compared with the widely applied model-free and model-based reinforcement learning (RL) algorithms, our Neural-PMP achieves higher sample-efficiency and performance in terms of control objectives. 
\end{abstract}


\section{Introduction}
Solving real-world optimal control problems is possessing huge engineering and societal values, ranging from running power grids, achieving robotics manipulation, and operating transportation networks. However, finding such real-world optimal controllers is often a challenging task, since real-world systems dynamics are usually unrevealed and complex due to their uniqueness, non-linearity and high-dimension~\cite{sutton1992reinforcement, jin2022learning}. Previous literature provides several extensively utilized approaches on building optimal controllers, including dynamic programming (DP) and model predictive control (MPC)~\cite{bellman1966dynamic}. However, most of these approaches require full knowledge of underlying state transitions, or use reduced-order models to facilitate numerical solution process for higher-dimensional problems. Such restrictions address difficulties on directly applying these approaches in solving complicated real-world optimal control problems with unknown system dynamics~\cite{lewis2012optimal, wang2009fast}.


Recent developments of Reinforcement Learning (RL) provides another direction to tackle this problem. Model-free deep RL algorithms relax the requirements over prior knowledge on system dynamics~\cite{Sutton1998}. Instead, agents can be trained to either learn the state-action value function (Q function) or learning the policies directly. 
Typical deep RL methods achieve impressive performances on certain control tasks~\cite{mnih2013playing, lillicrap2015continuous}. However, model-free deep RL methods often suffer from sample efficiency, requiring huge amount of trials and training time to fit effective policy networks. Moreover,  typical model-free RL lacks interpretability ~\cite{verma2018programmatically}, which raises reliability and safety concerns, especially for problems with hard constraints \cite{garcia2015comprehensive, 9198135}. 
To address these challenges, researchers revisit model-based approaches and develop model-based RL (MBRL). Such learning-based method usually first acquires a model of system dynamics and then find control policies based on learned models~\cite{606886}. 
Recent introduction of deep neural networks (NN) in data-driven control promises a versatile and efficient tool for controlling under unknown dynamics  \cite{8463189}. Such surrogate learned models are deployed in model-based RL (MBRL) frameworks~\cite{pmlr-v32-levine14, chen2019optimal}. Compared with model-free RL with only learning-based policy networks, NN dynamics in MBRL exhibit higher generalizability and effectiveness in unexplored state-action regions. The generalizability of NN dynamics helps significantly reducing the training samples for MBRL methods.
Yet typical MBRL algorithms such as  MPC with random shooting (RS-MPC) \cite{nagabandi2018neural} or policy networks  do not provide explicit mechanism for finding optimal decisions~\cite{dinev2022differentiable}. The lack of tractability severely influence the control performance and reliability of these MBRL algorithms, hence restricting their potential on real-world problems with unknown dynamics.


\begin{figure*}
  \begin{center}
    \includegraphics[width=13cm]{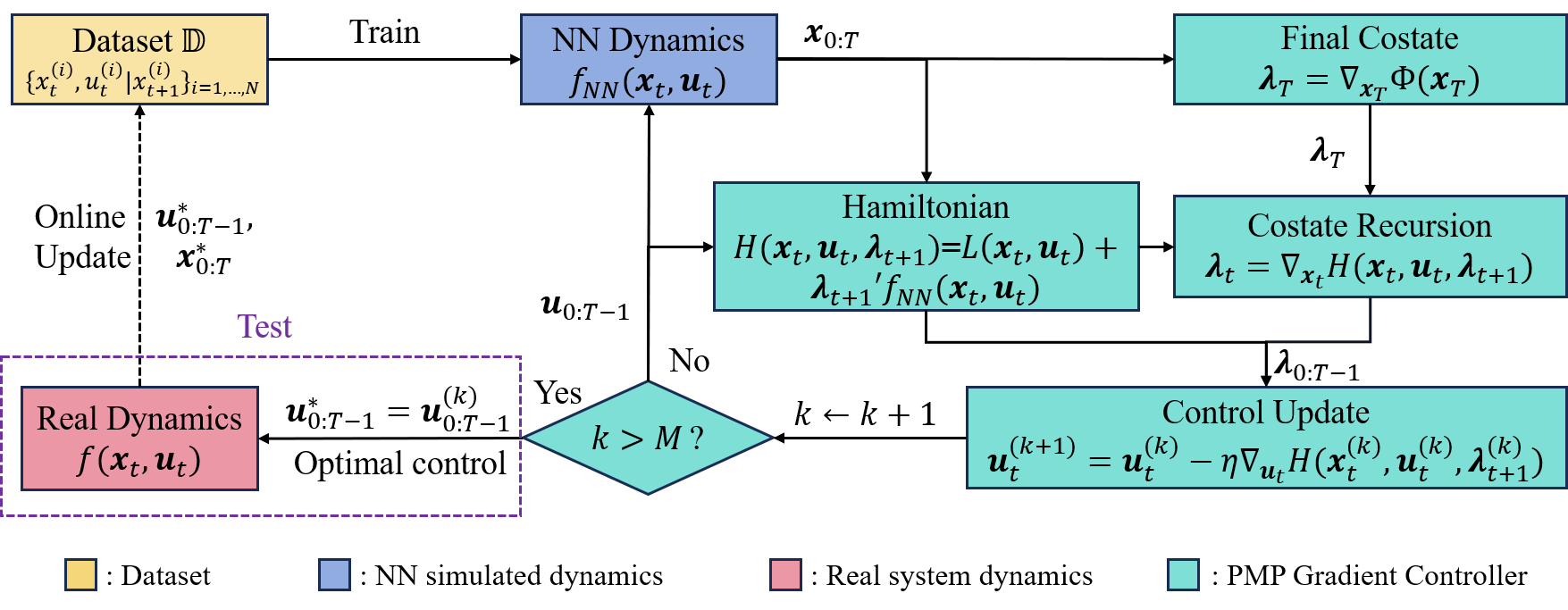}

  \end{center}
    \caption{Schematic of our proposed model learning and control framework based on PMP conditions. }     \label{fig:nnpmp}
\end{figure*}

In this paper, we design a novel and sample-efficient controller built upon accurate dynamics purely learned from data as well as theoretical foundations of Pontryagin's maximum principle (PMP). We leverage the potential of PMP into data-driven controller design with a fully differentiable architecture. PMP has been used to numerically solve the optimal control problems for a few decades~\cite{pontryagin1987mathematical}. It states a set of necessary conditions, including a minimum condition of the control Hamiltonian, for optimal control policies and optimal state trajectory. Unlike DP or MPC, solving PMP conditions can give the entire optimal solutions over the planning horizon. Working examples on control of electric vehicles~\cite{
chowdhury2020optimal}, supply chain~\cite{ivanov2012dynamic}, and financial markets~\cite{wang2010pontryagin} have been brought up. However, PMP's effectiveness relies on \emph{knowing the exact control Hamiltonian}, while it is less investigated for systems with unknown or unrevealed dynamics. To overcome such challenge, we seamlessly integrate learning  dynamics purely from collected state-action samples with NN into PMP-based optimal control. Once NN dynamics are well learned, we explicitly write out the Hamiltonian parameterized by the NN, making it possible to solve the derived optimality conditions using the surrogate NN model. 

Several recent works also considered data-driven design based on PMP principles. \cite{jin2020pontryagin, bottcher2022ai} introduce auxiliary models by differentiating PMP conditions, while the major objective for these methods are to help take gradients for policy network updates. In \cite{engin2023neural} authors directly apply PMP conditions and the Hamiltonian-Jacobian-Bellman (HJB) equations to guide the policy network training. Yet these approaches still involve sampling-expensive policy network fitting processes, and does not provide guarantee on the tractability of control actions.
Different from previous design of using policy network as controller, our major algorithm, Neural-PMP, directly establishes a differentiable procedure PMP-Gradient over PMP conditions to numerically solve the optimal control actions with learned NN system dynamics. Such iterative approach has been discussed in the literature, such as the continuous-time steepest gradient approach introduced in \cite{kirk2004optimal} for solving equations with PMP conditions involved. Neural-PMP differentiates from these methods in greatly expanding the applicability of PMP conditions under the scenarios with only data and NN model available.  We also prove with accurate NN model, PMP-Gradient controller can at least ensure convergence to local optimal solution, and under certain conditions can converge to global optimal.  Numerical simulations on a variety of control tasks including Linear Quadratic Regulator, charging of grid-connected lossy battery, control of single pendulum, and two MuJoCo locomotion tasks convincingly demonstrate our proposed Neural-PMP can: (1) outperform standard Linearized MPC models in control performance thanks to more accurate NN modeling; (2) achieve superior control performance with fewer training samples compared to deep RL algorithms, showcasing higher sample efficiency; (3) promote safer solutions with regard to state and action bounds versus MBRL using random search or model-free RL.

\vspace{-2pt}\section{Problem Formulation}
\subsection{Problem Statement}
In this paper, we consider  the discrete-time, finite horizon optimal control problem of a dynamical system. For a fixed time horizon $T$, we denote the system dynamics as
\begin{equation}
    \mathbf{x}_{t+1}=f(\mathbf{x}_{t}, \mathbf{u}_{t}), \quad \text{with given $\mathbf{x}_{0}$}.   
    \label{eqDynamics}
\end{equation}

Here $t=0,1,2,...,T$ is the time step within the given time horizon. 
$\mathbf{x}_{t}\in \mathbb{R}^m, \mathbf{u}_{t} \in \mathbb{R}^n$ denote the system state and the control variable at time step $t$ respectively. Eq. \eqref{eqDynamics} is termed as \textit{system dynamics}. In a classic optimal control problem, our goal is to find the optimal control policy $\mathbf{\pi} = \{\mathbf{u}_{0:T-1}^{*}\}$, which minimizes the \textit{cost functional} $J$ under the constraint of system dynamics and given initial state $\mathbf{x}_{0}$:
\begin{equation}
    \begin{aligned}
         \mathbf{\pi}=\arg \min_{\{\mathbf{u}_{0:T-1}\}} &J = \Phi(\mathbf{x}_{T}) + \sum_{t=0}^{T-1} L(\mathbf{x}_{t}, \mathbf{u}_{t})  \\
         \; s.t. \quad & \eqref{eqDynamics}, \quad 
         \mathbf{x}_{t} \in \mathcal{X}, \mathbf{u}_{t} \in \mathcal{U};
    \end{aligned}
    \label{eqOCProblem}
\end{equation}
where $\mathcal{X}\subseteq \mathbb{R}^n, \mathcal{U}\subseteq \mathbb{R}^m$ represent feasible regions for state and control variables respectively. $ L(\mathbf{x}_{t}, \mathbf{u}_{t})$  denotes \textit{running cost} for each time step, and  $\Phi(\mathbf{x}_{T})$ denotes the (optional) \textit{terminal cost} associated with terminal state $\mathbf{x}_{T}$. Without loss of generality, the terminal conditions can be treated either as a constraint or as a penalty term in $\Phi(\cdot)$. Also note $L(\cdot)$ and $\Phi(\cdot)$ are usually artificially designated functions.

In this work, we know the decision maker knows the exact function form of control objectives. Normally, with known, linear system dynamics, we can resort to 
off-the-shelf optimization solvers like convex optimization solver or MPC approaches to find the sequence of optimal actions $\mathbf{u}_{0:T-1}^{*}$. However, in many real-world cases, system dynamics are highly non-linear and unrevealed to controllers, such like  the robot locomotion models in robotic control tasks or battery chemical degradation models in the battery energy management tasks. Such situation makes these classic techniques less effective and not readily applicable. This motivates us to explore novel optimal control frameworks.


\vspace{-5pt}\subsection{Pontyragin Maximum Principle}
In this Subsection, we recall (discrete-time) Pontryagin’s Maximum Principle (PMP)~\cite{pontryagin1987mathematical}, which offers a set of necessary conditions that the optimal trajectory $\{\mathbf{x}_{0:T}^{*}, \mathbf{u}_{0:T-1}^{*}\}$ must satisfy in the optimal control problem Eq. \eqref{eqOCProblem}. To introduce these necessary conditions, we write the \textit{Hamiltonian}:
\begin{equation}
    H(\mathbf{x}_{t}, \mathbf{u}_{t}, \mathbf{\lambda}_{t+1}) = L(\mathbf{x}_{t}, \mathbf{u}_{t}) + \mathbf{\lambda}_{t+1}^{'} f(\mathbf{x}_{t}, \mathbf{u}_{t});
    \label{eqHam}
\end{equation}
where $\mathbf{\lambda}_{t}$ is named as \textit{costate} with $\mathbf{\lambda}_{t}'$ as its transpose. Based on PMP conditions, at time step $t$ in the horizon $[0, T-1]$ the optimal control $\mathbf{u}_{t}^{*}$ must satisfy the following conditions:
\begin{align}
    & \mathbf{\lambda}_{t}^{*} = \frac{\partial H(\mathbf{x}_{t}^{*}, \mathbf{u}_{t}^{*}, \mathbf{\lambda}_{t+1}^{*})}{\partial \mathbf{x}_{t}^{*}} 
    \label{eqCostateEquation};
    \\
    & \mathbf{\lambda}_{T}^{*} = \frac{\partial \Phi(\mathbf{x}_{T}^{*})}{\partial \mathbf{x}_{T}^{*}}
    \label{eqTerminal};
\\
    &\frac{\partial H(\mathbf{x}_{t}^{*}, \mathbf{u}_{t}^{*}, \mathbf{\lambda}_{t+1}^{*})}{\partial \mathbf{u}_{t}^{*}} = 0.
    \label{eqHamiltonMinimize}
\end{align}

{Eq. }\eqref{eqCostateEquation}-\eqref{eqHamiltonMinimize} are termed as \textit{PMP Conditions}. We refer to \cite{bryson2018applied} for the proof of PMP on standard discrete-time system 
\eqref{eqOCProblem}. Essentially, PMP conditions characterize the first-order conditions around the optimality for all kinds of systems. By directly solving PMP conditions {Eq. }\eqref{eqCostateEquation}-\eqref{eqHamiltonMinimize}, or integrating them into other algorithm blocks, we can overcome the failure of classical convex solver on complex non-linear systems.
However, this still requires full information of the control Hamiltonian along its first-order information with respect to both state variables and control variables. Yet in many real-world applications, $H(\mathbf{x}_{t}^{*}, \mathbf{u}_{t}^{*}, \mathbf{\lambda}_{t+1}^{*})$ may not be revealed to engineers or decision makers, or only partial of the dynamics information are available. Moreover, previous attempts on using PMP conditions for finding closed-form solutions of the optimal controller only work with limited form of dynamics or cost functions~\cite{chowdhury2020optimal}.  This motivates us to examine how to integrate data-driven dynamical model into PMP conditions, and investigate the computational method to solve for $\mathbf{u}_{t}^{*}$. 

\section{Model-based Optimal Controller Design}
In this section, we demonstrate our proposed model-based approach for optimal control named \textbf{Neural-PMP}, which is composed of two major parts: i). train neural networks (NN) to learn system dynamics offline from data; and ii). use a PMP-based gradient method to obtain the optimal control policy based on the learned dynamics. Proposed scheme of Neural-PMP is illustrated in Fig. \ref{fig:nnpmp}.

\subsection{System Identification with Neural Networks}
\label{sec:NNPMP}
To leverage PMP's advantages on finding optimal control sequences of unknown dynamics, we first need an effective method to model such dynamics from collected data. By doing so, we are able to formulate the required control Hamiltonian explicitly. 
In our proposed approach, we utilize the function approximation capabilities of NN, and parameterize the system dynamics either wholly or partially as NN: 
\begin{equation}
    \mathbf{x}_{t+1}=f_{NN}(\mathbf{x}_{t}, \mathbf{u}_{t};\theta) \quad \text{with given $\mathbf{x}_{0}$};
    \label{equ:NN_dynamics}
\end{equation}
Usually we train the NN-simulated system dynamics $f_{NN}$ offline with dataset $\mathbb{D}=\{\mathbf{x}_{t}^{(i)}, \mathbf{u}_{t}^{(i)}| \mathbf{x}_{t+1}^{(i)}  \}_{i=1,2,...,N}$. To establish $\mathbb{D}$, we randomly collect state-control pairs $\{\mathbf{x}_{t}^{(i)}, \mathbf{u}_{t}^{(i)}\}$ from state and control space $\mathcal{X}$, $\mathcal{U}$, and use ground-truth system dynamics $f$ to generate next state $\mathbf{x}_{t+1}^{(i)}$ as supervised labels. Once the training data is collected, we learn parameters $\theta$ of $f_{NN}$ by minimizing mean squared error (MSE):
\begin{equation}
    \min_{\theta} \; \frac{1}{N}\sum_{i=1}^{N} (f_{NN}(\mathbf{x}_{t}^{(i)},\mathbf{u}_{t}^{(i)};\theta) - \mathbf{x}_{t+1}^{(i)})^{2}.
    \label{equ:lossSys}
\end{equation}

To improve the dynamics approximation accuracy, we can also train $f_{NN}$ in an online fashion by updating $\mathbb{D}$ with new state-control pairs executed in the control process and re-training $f_{NN}$, which is also broadly used in RL community. 
Based on the trained system dynamics and designated optimizing objective, we formulate the Hamiltonian and establish equations for optimal control and state trajectory with PMP Conditions {Eq. }\eqref{eqCostateEquation}-\eqref{eqHamiltonMinimize}. However, as non-linearity shows up in NN-simulated system dynamics $f_{NN}$, we still face challenges in directly solving these equations, which motivates us to find numerical way to solve these equations with learned dynamics involved.
\vspace{-10pt}
\subsection{PMP-Gradient Controller Design}
\label{sec:Controller}
With accurately learned function $f_{NN}$ in Eq. \eqref{equ:NN_dynamics}, we are able to establish control Hamiltonian $H(\mathbf{x}_{t}, \mathbf{u}_{t}, \mathbf{\lambda}_{t+1}) = L(\mathbf{x}_{t}, \mathbf{u}_{t}) + \mathbf{\lambda}_{t+1}^{'} f_{NN}(\mathbf{x}_{t}, \mathbf{u}_{t};\theta)$, as well as PMP conditions  Eq. \eqref{eqCostateEquation}-\eqref{eqHamiltonMinimize} for the system. While the ultimate goal is to find solutions for these equations, which help us get the optimal control policy $\pi=\mathbf{u}_{0:T-1}^{*}$ of the sequential decision-making problems. Due to the non-linearity and complicated function space of $f_{NN}$, it is challenging to directly obtain solutions in an analytical way. Algorithms such as random shooting has tractability and safety concerns, as there is no guarantee on action's performance, while the resulting dynamics are not regulated.   To overcome these difficulties, we are inspired by the iterative approaches and design gradient-based method PMP-Gradient to compute numerical solutions for $\pi$.

In this approach, we start from an initialized random control policy $\mathbf{u}^{(1)}_{0:T-1}$ with superscript denoting action updating iterations. And we use such control to retrieve corresponding state sequence $\mathbf{x}^{(1)}_{0:T}$ based on NN-simulated system dynamics Eq. \eqref{equ:NN_dynamics}. Then based on it, we recursively compute costate  $\mathbf{\lambda}^{(1)}_{0:T}$ with PMP Conditions Eq. \eqref{eqCostateEquation}-\eqref{eqTerminal}:
\begin{equation}
    \begin{aligned}
        &\mathbf{\lambda}_{T}^{(1)} = \frac{\partial \Phi(\mathbf{x}_{T}^{(1)})}{\partial \mathbf{x}_{T}^{(1)}}; \\
        &\mathbf{\lambda}_{t}^{(1)} = \frac{\partial H(\mathbf{x}_{t}^{(1)}, \mathbf{u}_{t}^{(1)}, \mathbf{\lambda}_{t+1}^{(1)})}{\partial \mathbf{x}_{t}^{(1)}}. \vspace{-10pt}
    \end{aligned}
\end{equation}

Based on values of obtained states and costates, we update controls by taking gradients of Hamiltonian $H(\mathbf{x}^{(1)}_{t}, \mathbf{u}^{(1)}_{t}, \mathbf{\lambda}^{(1)}_{t+1})$ \textit{w.r.t.} $\mathbf{u}^{(1)}_{t}$ and apply a gradient descent optimizer with learning rate $\eta$ to update $\mathbf{u}^{(2)}_{t}$:
\begin{equation}
\label{equ:u_iter}
    \mathbf{u}^{(2)}_{t} = \mathbf{u}^{(1)}_{t} - \eta \frac{\partial H(\mathbf{x}_{t}^{(1)}, \mathbf{u}_{t}^{(1)}, \mathbf{\lambda}_{t+1}^{(1)})}{\partial \mathbf{u}_{t}^{(1)}}, \; t \in [0, T].
\end{equation}

By repeating the above process until maximum iterations $M$, we iteratively update control policy $\mathbf{u}_{0:T-1}$ to reach solutions of PMP Condition Eq. \eqref{eqHamiltonMinimize}. The full algorithm is illustrated in Algorithm \ref{al1}.


\begin{algorithm}[]
\small
    \caption{Neural-PMP}
    \label{al1}
    \begin{algorithmic}[1]
        \Require Initial state $\mathbf{x}_{0}$; Maximum iterations $M$; Learning rate $\eta$; Dataset $\mathbb{D}$ for training NN-simulated system dynamics $ \mathbf{x}_{t+1}=f_{NN}(\mathbf{x}_{t}, \mathbf{u}_{t};\theta)$;
         \Require {Running cost  $L(\mathbf{x}_{t}, \mathbf{u}_{t})$; Terminal cost $\Phi(\mathbf{x}_{T})$.}
        \State {$k=1$.}
        \State Initialize control policy $\mathbf{u}_{0:T-1}^{(1)}$.
        \State Learn $\theta$ for NN dynamics $f_{NN}$ with $\mathbb{D}$.
        \While{$k \leq M$}
        \State \quad Establish form of Hamiltonian: $H(\mathbf{x}_{t}, \mathbf{u}_{t}, \mathbf{\lambda}_{t+1}) = L(\mathbf{x}_{t}, \mathbf{u}_{t}) + \mathbf{\lambda}_{t+1}^{'} f_{NN}(\mathbf{x}_{t}, \mathbf{u}_{t};\theta)$
        \State \quad Compute $\mathbf{x}_{0:T}^{(k)}$ with current control policy $\mathbf{u}_{0:T-1}^{(k)}$, based on $ \mathbf{x}_{t+1}=f_{NN}(\mathbf{x}_{t}, \mathbf{u}_{t})$ and $\mathbf{x}_{0}$. 
        \State \quad Compute terminal costate $\mathbf{\lambda}_{T}^{(k)}$ by $\mathbf{\lambda}_{T}^{(k)} = \frac{\partial \Phi(\mathbf{x}_{T}^{(k)})}{\partial \mathbf{x}_{T}^{(k)}}$.
        \State \quad Compute costate $\mathbf{\lambda}_{t}^{(k)}$ for $t=0, 1,..,T-1$, based on $\mathbf{\lambda}_{t}^{(k)} = \frac{\partial H(\mathbf{x}_{t}^{(k)}, \mathbf{u}_{t}^{(k)}, \mathbf{\lambda}_{t+1}^{(k)})}{\partial \mathbf{x}_{t}^{(k)}}$ and $\mathbf{\lambda}_{T}^{(k)}$.
        \State \quad Update Control Policy: $\mathbf{u}_{t}^{(k+1)} \leftarrow \mathbf{u}_{t}^{(k)} - \eta  \frac{\partial H(\mathbf{x}_{t}^{(k)}, \mathbf{u}_{t}^{(k)}, \mathbf{\lambda}_{t+1}^{(k)})}{\partial \mathbf{u}_{t}^{(k)}}$.
        \State \quad (If apply online training) 
        \quad Apply $\mathbf{u}_{0:T-1}^{(k)}$ to real system $f$, collect state-action pairs $\{\mathbf{x}_{t}^{(k)}, \mathbf{u}_{t}^{(k)}|f(\mathbf{x}_{t}^{(k)}, \mathbf{u}_{t}^{(k)})\}_{t=0:T-1}$, add them to $\mathbb{D}$ and re-train $f_{NN}$.
        \State \quad $k \leftarrow k+1$
        \EndWhile
        \State $\mathbf{u}_{0:T-1}^{*} \leftarrow \mathbf{u}_{0:T-1}^{(k)}$
    \end{algorithmic}
\end{algorithm}

\begin{table*}
\scriptsize
\centering
\begin{tabular}{c|ccccc} 
\hline
             & $T$ & $(m, n)$ & \textbf{Running Cost} $L(\mathbf{x}_{t}, \mathbf{u}_{t})$ & \textbf{Terminal Cost} $\Phi(\mathbf{x}_{T})$ &  \textbf{System Dynamics} $f(\mathbf{x}_{t}, \mathbf{u}_{t})$ \\ 
\hline
LQR & 10 & (5, 3) &  $\mathbf{x}_{t}^{'}\mathbf{Q}\mathbf{x}_{t}+\mathbf{u}_{t}^{'}\mathbf{R}\mathbf{u}_{t}$& $\mathbf{x}_{T}^{'}\mathbf{Q_T}\mathbf{x}_{T}$ &  $\mathbf{A}\mathbf{x}_{t} + \mathbf{B}\mathbf{u}_{t}$  \\ 
\hline
Battery     & 24 & (1, 1) & $p_t\mathbf{u}_t + \alpha \mathbf{u}^{2}_t + P_e(\mathbf{x}_t)$ & $\gamma(\mathbf{x}_T-\mathbf{x}_f)^{2}$ &  $\zeta(\mathbf{u}_t)\mathbf{u}_t + \mathbf{x}_t$ \\ 
\hline
Pendulum    & 10 & (2, 1) & $w_q(q_t-q_f)^2+w_{dq}(dq_t-dq_f)^2+w_u\mathbf{u}_t^2$ & $w_q(q_T-q_f)^{2}+w_dq(dq_T-dq_f)^{2}$ & $\mathbf{x}_t + \Delta [dq_t, \frac{u_t-mglq_t-\sigma sin(q_t)}{I}]^{'}$  \\ 
\hline
Swimmer       & 500 & (10, 2) &$w_c\mathbf{u}_t^{'}\mathbf{u}_t$  & $w_f(\mathbf{x}_{0}[0]-\mathbf{x}_T[0])$ &  unknown \\ 
\hline
HalfCheetah & 500 & (18, 6) & $w_c\mathbf{u}_t^{'}\mathbf{u}_t$ & $w_f(\mathbf{x}_{0}[0]-\mathbf{x}_T[0])$ &  unknown \\
\hline
\end{tabular}
\caption{Settings of testing environments. For Swimmer and HalfCheetah we do not know the exact formulation of the system dynamics, and all algorithms need to learn the dynamics or policy.}
\label{tab:tasks}
\end{table*}


Essentially, we utilize a fully differentiable procedure by using the learned dynamics $f_{NN}$ with the corresponding Hamiltonian and the terminal cost. Unlike typical policy network updates in model-free RL, the inference of control actions in our Neural-PMP is training-free, which further reduce computation need during training. Meanwhile, when $f_{NN}$ approximates ground-truth $f$ accurately, we have the following Lemma regarding step $5-6$ of Neural-PMP Algorithm \ref{al1} and Theorem for cost $J$ in w.r.t Eq. \eqref{eqOCProblem}:

\begin{prop}
\label{prop1}
Implementing  Eq. \eqref{equ:u_iter} is equivalent to taking gradient descent step $\mathbf{u}_{t}^{(k+1)} =\mathbf{u}_{t}^{(k)}- \eta \frac{\partial J}{ \partial \mathbf{u}_{t}^{(k)}}$.
\end{prop}

\begin{theorem}
\label{theorem1}
\quad Suppose Objective $J$ is  \textit{differentiable and strictly convex} with respect to $\mathbf{u}_{t}^{(k)}$, and $\frac{\partial J}{ \partial \mathbf{u}_{t}^{(k)}}$ is \textit{Lipschitz Continuous} with constant $L>0$. Then gradient descent $\mathbf{u}_{t}^{(k+1)} =\mathbf{u}_{t}^{(k)}- \eta \frac{\partial J}{ \partial \mathbf{u}_{t}^{(k)}}$ guarantees global optimal with learning rate $\eta \leq 1/L$.
\end{theorem}

Based on Lemma \ref{prop1} and Theorem \ref{theorem1}, we can infer that our PMP-Gradient Controller has global convergence guarantees with convergence rate $\mathit{O}(1/\eta)$ when $J$ is a convex function \textit{w.r.t.} control variable $\mathbf{u}_{t}$ for all $t=0, 1,..,T-1$ (and has at least local convergence guarantees for general cases). The proof of Lemma \ref{prop1} and Theorem \ref{theorem1} can be found in Appendix. \ref{ConvergenceProof} in our paper's online version~\cite{gu2022pontryagin}. These observations demonstrate the generalizability of our Neural-PMP based on accurately trained NN dynamics across the entire state-action space. Indeed, our approach is more sample efficient as fitting dynamics is much more efficient than training policy networks in model-free RL~\cite{nagabandi2018neural}. To accelerate policy convergence, other gradient-based optimizers like conjugate gradient (CG) or Newton's method can be also applied.


Different from directly performing gradient descent over objective $J$, PMP-Gradient controller does not need to consider terms with future states $\mathbf{x}_{t+1:T}$ when taking derivatives of $H(\mathbf{x}_{t}, \mathbf{u}_{t}, \mathbf{\lambda}_{t+1})$ \textit{w.r.t.} $\mathbf{u}_{t}$. This helps us eliminate computations of Hessian matrices $\frac{\partial \mathbf{x}_{t+1}}{\partial \mathbf{x}_{t}}$, hence reducing the size of computation graph for calculating gradients and greatly improving the computation time. For problems with hard state and action bounds, we modify objective and step 7 in Algorithm \ref{al1}. Detailed settings can be found in Appendix B in our paper's online version~\cite{gu2022pontryagin}. With this design, we can mitigate the risk of state and action exceeding bounds, as this controller avoids any uncontrollable random elements.

\vspace{-10pt}\section{Numerical Examples}
\label{sec:simulations}
\begin{table*}[]
\centering

\begin{tabular}{c|x{2.5cm} cx{2.5cm} cx{2.5cm} cx{2.5cm}cx{2.5cm} cx{2.5cm}} 
\hline
              & LQR (C) & Battery (C) & Pendulum (C) & Swimmer (R) & HalfCheetah (R) \\ 
\hline
Linearized  &       13.43$\pm$0.00       &      74581.94$\pm$13446.32 &          1081.97$\pm$0.00           &      -7.61$\pm$0.37               &       -261.44$\pm$16.45     \\
\hline
PPO  &       82.48$\pm$23.12       &      19619.10$\pm$24318.00           &          1086.11$\pm$18.14            &      34.99$\pm$0.85               &       96.73$\pm$47.00                   \\ 
\hline
RS-MPC &     552.31$\pm$96.98         &         300.23$\pm$206.92         &       1129.46$\pm$23.23            &      26.81$\pm$2.72            &              -47.62$\pm$44.77         \\ 
\hline
\textbf{Ours} &      13.53$\pm$0.09        &      -3.20$\pm$0.00               &     993.53$\pm$48.08                 &     31.12$\pm$1.85               &     313.77$\pm$131.96                    \\ 
\hline
Number of Samples       & 2,000         & 2,000             & 2,000              & 100,000           & 100,000                \\
\hline
\end{tabular}
\caption{Performances under varying number of training samples. The performances are averaged for 10 runs test.}
\label{tab:results}
\end{table*} 

\begin{table}[]
\centering
\begin{tabular}{c|c|c|c|c} 
\hline
       & \textbf{Linearized}       & \textbf{PPO} & \textbf{RS-MPC} & \textbf{Ours}  \\ 
\hline
\textbf{SBER} & 54.3\%$\pm$2.3\% & 33.7\%$\pm$21.3\%            & 5.1\%$\pm$1.4\%               & 0.0\%$\pm$0.0\%                            \\
\hline
\end{tabular}
\caption{State Bounds Exceeding Rates (SBER) on the Battery simulation. Results are averaged for 3 runs (each contains 10 trials).\vspace{-10pt} }
\label{tab:exceed}
\end{table}

In this section, we evaluate the optimal control performance of our proposed Neural-PMP controller and illustrate its versatile capabilities on numerical simulations of 5 tasks. In simulations we benchmark against: i). a Linear-Convex solver, which we firstly use linear regression to approximate a linearized system dynamics and then solve  MPC problem using cvxpy solver;
ii). state-of-the-art model-free RL algorithm Proximal Policy Optimization (PPO); and iii). model-based random shooting MPC (RS-MPC) controller (based on NN-simulated system dynamics)~\cite{nagabandi2018neural}. To benchmark performances, for all simulations we use a Google Colab T4 GPU with 12GB RAM.

\vspace{-5pt}\subsection{Simulated Control Tasks}
\label{exp:Tasks}
The settings of 5 benchmarked control tasks (including time horizon $T$, state and action dimension $m, n$, running cost $ L(\mathbf{x}_{t}, \mathbf{u}_{t})$, terminal cost $\Phi(\mathbf{x}_{T})$ and system dynamics $f(\mathbf{x}_{t}, \mathbf{u}_{t})$  are illustrated in Table. \ref{tab:tasks}. We note that in all test cases, it is  assumed that the ground truth dynamics are not revealed to the controller. Detailed settings of these systems can be found in Appendix C in our paper's online version~\cite{gu2022pontryagin}. Our codes for the controller and numerical simulations are publicly available at \url{https://github.com/ChengyangGU/NeuralPMP2024}.

\begin{figure}[]
    \centering
    \includegraphics[width=7.5cm]{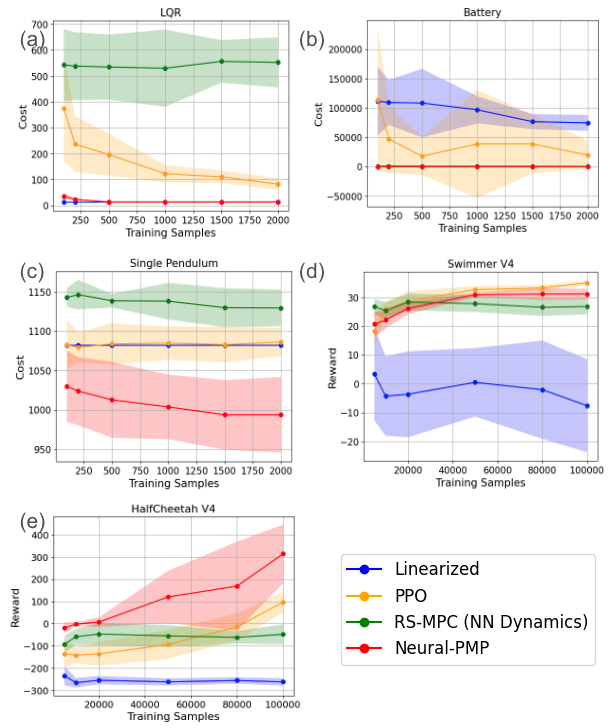}
    \caption{Sample-Performance (Cost/Reward) curves on (a) LQR (C); (b) Battery (C); (c) Pendulum (C); (d) Swimmer (R);  (e) HalfCheetah (R).\vspace{-25pt}}
    \label{fig:mainfig}
\end{figure}

\noindent \textbf{1. Linear Quadratic Regulator (LQR)}. In this case, we set up a  finite-horizon Linear Quadratic Regulator (LQR) model with 5 states and 3 actions with choice of stablizable $(A,\,B)$.  

\noindent \textbf{2. Battery}. We consider optimal arbitrage of a battery. The battery maximizes profits via participating in electricity markets and making charging or discharging decisions under time-varying prices. A profitable energy storage arbitrage plan is crucial, since it encourages utilization of stochastic renewables and addressing energy sustainability \cite{dell2001energy}. The battery has a nonlinear and unknown charging efficiency with respect to battery's state-of-charge.


\noindent \textbf{3. Pendulum}. We consider controlling a single pendulum system, where control variable $\mathbf{u}_{t}$ denotes the applied torque, while the 2-dimension state $[q_t, dq_t]^{'}$ include angular velocity $q_t$ and angular acceleration $dq_t$.

\noindent \textbf{4. MuJoCo Swimmer}. We consider the control of MuJoCo robot Swimmer \cite{todorov2012mujoco}. The problem involves a 2-dimension control variable $\mathbf{u}_t$ and a 10-dimension state $\mathbf{x}_t$. The first element of state $\mathbf{x}_t[0]$ denotes the robot's x-coordinate. For optimal control, we want the Swimmer robot to move forward along x-axis as much as possible with small control cost.  $w_c$ and $w_f$ are cost weights. 

\noindent \textbf{5. MuJoCo HalfCheetah}. This problem involves a 6-dimension control variable $\mathbf{u}_t$ and a 18-dimension state $\mathbf{x}_t$. The control objective is same as Swimmer.

\vspace{-10pt}\subsection{Numerical Results}
We evaluate the control performance, sample-efficiency and safety of our proposed Neural-PMP on the 5 tasks mentioned in Section \ref{exp:Tasks}. For LQR, Single Pendulum, Swimmer and HalfCheetah simulations, we apply neural network $f_{NN}$ to learn the full system dynamics $f(\mathbf{x}_{t}, \mathbf{u}_{t})$, while in Battery case we use $f_{NN}$ to learn part of the dynamics: the charging efficiency function $\zeta(\mathbf{u}_t)$. In all 5 simulations, we compare our Neural-PMP method against Linearized model, classic model-free RL method PPO and model-based RL method RS-MPC (based on the same NN-simulated system dynamics in our Neural-PMP method). For all environments tested we find Neural-PMP converges during policy inference.


Table. \ref{tab:results} shows the control performance of each control method across all 5 environments. In general, the results demonstrate that our Neural-PMP outperforms Linearized model, PPO, and RS-MPC in terms of control performance using same number of training samples. For example, in Single Pendulum simulation, with NN-dynamics trained by 2,000 samples, our Neural-PMP reduces the cost by at least 90 compared with Linearized model, PPO and RS-MPC, yielding more than 8.2\% improvement; while in HalfCheetah case, with NN-dynamics trained by 100,000 samples, our method shows more than 200 leading in reward, which is huge improvement compared to the best result $96.73$ (achieved by PPO) among the rest 3 methods. We also note that in specific cases with strong local optima like Swimmer \cite{franceschetti2022making}, though Neural-PMP achieves competitive control performance, it can converge to local optima from some initialized actions,  making it underperform model-free PPO with a very narrow gap. We will further address these concerns in the future work.  



\begin{figure}[hbt]
    \centering
    \includegraphics[width=7.5cm]{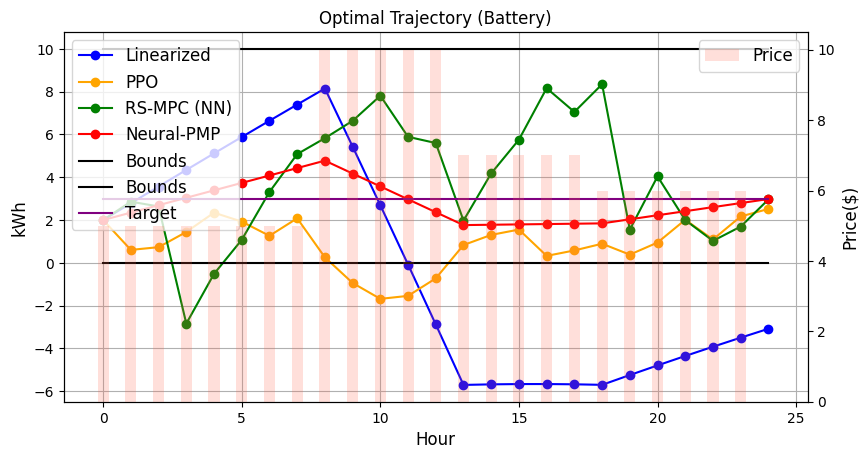}
    \caption{Battery state trajectory obtained by Neural-PMP is the only one not exceeding state bounds. \vspace{-15pt}}
    \label{fig:Trajectory}
\end{figure}

To validate the sample efficiency, Fig. \ref{fig:mainfig} demonstrates that our method maintains a significant advantage in control performance with varying number of training data. In most cases, even with substantially fewer training samples for learning the system dynamics, our method achieves higher rewards/ lower costs. Furthermore, simulation results highlight advantages of our Neural-PMP on satisfying safety constraints defined by the state or action constraints. Table. \ref{tab:exceed} shows the average State Bounds Exceeding Rate (SBER) in battery simulation with 2,000 training samples, which records the proportion of state trajectories that exceed bounds obtained by various controllers across total time steps. Fig. \ref{fig:Trajectory} shows one-day state trajectory comparison for the Battery case. We can observe only our method constrains the battery state-of-charge within the pre-defined $0-10$ kWh limits, while other algorithms' resulting actions cause the battery to be easily overdischarged with state-of-charge smaller than $0$ kWh. More importantly, unlike the irregular solution obtained by PPO or RS-MPC, our proposed algorithm can charge the battery when price is lower, while selling power at hours with higher price during noon, which is more interpretable. These results further indicate that our Neural-PMP yields safer and realistic solutions, and hence improves the control performance significantly. In summary, our Neural-PMP outperforms other benchmarked methods in terms of control performance and sample efficiency across all 5 environments, and can achieve safer solutions. These advantages of our Neural-PMP is attributed to its ability to learn the unrevealed dynamics while utilizing PMP conditions for finding optimal control sequences.


\vspace{-8pt}\section{Conclusion and Discussion}
In this paper, we showed that the classic  Pontryagin's Principle can be seamlessly integrated into data-driven settings. The proposed Neural-PMP is able to both accurately simulate and optimally control unknown dynamical systems. Simulations on five numerical examples demonstrated our Neural-PMP outperforms standard optimization-based solver using linearized dynamics, model-free  and model-based RL under a variety of settings. For complex real-world systems, the proposed method is a more sample-efficient, generalizable and safer optimal controller. For future work, we will investigate the effects of NN modeling errors and initialization of policies in our Neural-PMP controller design. 
We aim to address these challenges and seek theoretical insights to enhance our Neural-PMP as a more robust and efficient model-based optimal controller.\vspace{-10pt}

\bibliographystyle{IEEEtran}
\bibliography{bib}

\newpage
\appendix
\label{appendix}
\subsection{Convergence Proof of PMP-Gradient Controller}
\label{ConvergenceProof}

\noindent \textit{Lemma 1.} \quad Implementing  Eq. \eqref{equ:u_iter} is equivalent to taking gradient descent step $\mathbf{u}_{t}^{(k+1)} =\mathbf{u}_{t}^{(k)}- \eta \frac{\partial J}{ \partial \mathbf{u}_{t}^{(k)}}$.


\begin{proof}
    Recall the formulation of Hamiltonian $H(\mathbf{x}_{t}, \mathbf{u}_{t}, \mathbf{\lambda}_{t+1})$ is:
\begin{equation}
    H(\mathbf{x}_{t}, \mathbf{u}_{t}, \mathbf{\lambda}_{t+1}) = L(\mathbf{x}_{t}, \mathbf{u}_{t}) + \mathbf{\lambda}_{t+1}^{'} f(\mathbf{x}_{t}, \mathbf{u}_{t});
    \label{eq:appendixH}
\end{equation}
Based on Eq. (\ref{eq:appendixH}) and PMP Conditions we get:
\begin{equation}
    \begin{aligned}
        &\mathbf{\lambda}_{T}^{(k)} = \frac{\partial \Phi(\mathbf{x}_{T}^{(k)})}{\partial \mathbf{x}_{T}^{(k)}} \\
        &\mathbf{\lambda}_{t}^{(k)} = \frac{\partial H(\mathbf{x}_{t}^{(k)}, \mathbf{u}_{t}^{(k)}, \mathbf{\lambda}_{t+1}^{(k)})}{\partial \mathbf{x}_{t}^{(k)}} \\ 
        & =  \frac{\partial L(\mathbf{x}_{t}^{(k)}, \mathbf{u}_{t}^{(k)})}{\partial \mathbf{x}_{t}^{(k)}} + \mathbf{\lambda}_{t+1}^{(k)} \frac{\partial f(\mathbf{x}_{t}^{(k)}, \mathbf{u}_{t}^{(k)})}{\partial \mathbf{x}_{t}^{(k)}}
    \end{aligned}
    \label{eq10}
\end{equation}
Notice that:
\begin{equation}
    f(\mathbf{x}_{t}^{(k)}, \mathbf{u}_{t}^{(k)}) = \mathbf{x}_{t+1}^{(k)}
\end{equation}
Hence, we have:
\begin{equation}
    \begin{aligned}
        \mathbf{\lambda}_{T-1}^{(k)} &= \frac{\partial L(\mathbf{x}_{T-1}^{(k)}, \mathbf{u}_{T-1}^{(k)})}{\partial \mathbf{x}_{T-1}^{(k)}} + \mathbf{\lambda}_{T}^{(k)} \frac{\partial \mathbf{x}_{T}^{(k)}}{\partial \mathbf{x}_{T-1}^{(k)}} \\
        &= \frac{\partial L(\mathbf{x}_{T-1}^{(k)}, \mathbf{u}_{T-1}^{(k)})}{\partial \mathbf{x}_{T-1}^{(k)}} + \frac{\partial \Phi(\mathbf{x}_{T}^{(k)})}{\partial \mathbf{x}_{T}^{(k)}} \frac{\partial \mathbf{x}_{T}^{(k)}}{\partial \mathbf{x}_{T-1}^{(k)}} \\
        &=  \frac{\partial L(\mathbf{x}_{T-1}^{(k)}, \mathbf{u}_{T-1}^{(k)})}{\partial \mathbf{x}_{T-1}^{(k)}} + \frac{\partial \Phi(\mathbf{x}_{T}^{(k)})}{\partial \mathbf{x}_{T-1}^{(k)}}
    \end{aligned}
    \label{eq:caseT1}
\end{equation}
Based on Eq. (\ref{eq10}) and Eq. (\ref{eq:caseT1}) do the recursion:
\begin{equation}
     \mathbf{\lambda}_{t}^{(k)}=\left\{
\begin{aligned}
&\sum_{i=t}^{T-1}(\frac{\partial L(\mathbf{x}_{i}^{(k)}, \mathbf{u}_{i}^{(k)})}{\partial \mathbf{x}_{t}^{(k)}}) + \frac{\partial \Phi(\mathbf{x}_{T}^{(k)})}{\partial \mathbf{x}_{t}^{(k)}}   &0 \leq t \leq T-1, \\
&\frac{\partial \Phi(\mathbf{x}_{T}^{(k)})}{\partial \mathbf{x}_{t}^{(k)}}   & t=T. \\
\end{aligned}
\right.
\label{eqCostateForm}
\end{equation}
Based on Eq. (\ref{eqCostateForm}) we compute $\frac{\partial H}{\partial \mathbf{u}_{t}^{(k)}}$ ($0 \leq t \leq T-1$):
\begin{equation}
    \begin{aligned}
        &\frac{\partial H}{\partial \mathbf{u}_{t}^{(k)}}  = \frac{\partial L(\mathbf{x}_{t}^{(k)}, \mathbf{u}_{t}^{(k)})}{\partial \mathbf{u}_{t}^{(k)}} + 
        \mathbf{\lambda}_{t+1}^{(k)} \frac{\partial \mathbf{x}_{t+1}^{(k)}}{\partial \mathbf{u}_{t}^{(k)}} \\
       & = \frac{\partial L(\mathbf{x}_{t}^{(k)}, \mathbf{u}_{t}^{(k)})}{\partial \mathbf{u}_{t}^{(k)}} +\sum_{i=t+1}^{T-1}(\frac{\partial L(\mathbf{x}_{i}^{(k)}, \mathbf{u}_{i}^{(k)})}{\partial \mathbf{u}_{t}^{(k)}}) + \frac{\partial \Phi(\mathbf{x}_{T}^{(k)})}{\partial \mathbf{u}_{t}^{(k)}} \\
        & =\sum_{i=t}^{T-1}(\frac{\partial L(\mathbf{x}_{i}^{(k)}, \mathbf{u}_{i}^{(k)})}{\partial \mathbf{u}_{t}^{(k)}}) +  \frac{\partial \Phi(\mathbf{x}_{T}^{(k)})}{\partial \mathbf{u}_{t}^{(k)}}
    \end{aligned}
    \label{eqHu}
\end{equation}
According to Eq. (\ref{eqOCProblem}) we compute $\frac{\partial J}{ \partial \mathbf{u}_{t}^{(k)}}$:
\begin{equation}
    \frac{\partial J}{ \partial \mathbf{u}_{t}^{(k)}} = \frac{\partial ( \sum_{t=0}^{T-1} L(\mathbf{x}_{t}^{(k)}, \mathbf{u}_{t}^{(k)}) +  \Phi(\mathbf{x}_{T}^{(k)}))}{\partial \mathbf{u}_{t}^{(k)}}
\end{equation}
Notice that:
\begin{equation}
    \frac{\partial ( \sum_{i=0}^{t-1} L(\mathbf{x}_{i}^{(k)}, \mathbf{u}_{i}^{(k)})}{\partial \mathbf{u}_{t}^{(k)}} = 0
\end{equation}
Hence we have
\begin{equation}
     \frac{\partial J}{ \partial \mathbf{u}_{t}^{(k)}} = \sum_{i=t}^{T-1}(\frac{\partial L(\mathbf{x}_{i}^{(k)}, \mathbf{u}_{i}^{(k)})}{\partial \mathbf{u}_{t}^{(k)}}) +  \frac{\partial \Phi(\mathbf{x}_{T}^{(k)})}{\partial \mathbf{u}_{t}^{(k)}}
     \label{eqJu}
\end{equation}
Compare Eq. (\ref{eqHu}) with Eq. (\ref{eqJu}), we have:
\begin{equation}
    \frac{\partial J}{ \partial \mathbf{u}_{t}^{(k)}} = \frac{\partial H}{ \partial \mathbf{u}_{t}^{(k)}}
\end{equation}
Hence, PMP Gradient Controller $\mathbf{u}_{t}^{(k+1)} \leftarrow \mathbf{u}_{t}^{(k+1)} - \eta  \frac{\partial H(\mathbf{x}_{t}^{(k)}, \mathbf{u}_{t}^{(k)}, \mathbf{\lambda}_{t+1}^{(k)})}{\partial \mathbf{x}_{t}^{(k)}}$ is equivalent to gradient descent $\mathbf{u}_{t}^{(k+1)} =\mathbf{u}_{t}^{(k)}- \eta \frac{\partial J}{ \partial \mathbf{u}_{t}^{(k)}}$.
\end{proof}

\vspace{+15pt}
\noindent \textit{Theorem 2.} \quad Suppose Objective $J$ is a \textit{differentiable and convex function} \textit{w.r.t} $\mathbf{u}_{t}^{(k)}$, and $\frac{\partial J}{ \partial \mathbf{u}_{t}^{(k)}}$ is \textit{Lipschitz Continuous} with constant $L>0$. Then gradient descent $\mathbf{u}_{t}^{(k+1)} =\mathbf{u}_{t}^{(k)}- \eta \frac{\partial J}{ \partial \mathbf{u}_{t}^{(k)}}$ guarantees global optimal with learning rate $\eta \leq 1/L$.
\vspace{+3pt}

\begin{proof}
Since $\frac{\partial J}{ \partial \mathbf{u}_{t}^{(k)}}$ is \textit{Lipschitz Continuous}, we have:
\begin{equation}
    J(\mathbf{y}) = J(\mathbf{z}) + \frac{\partial J}{\partial \mathbf{z}}(\mathbf{y}-\mathbf{z}) + \frac{1}{2}L\Vert\mathbf{y}-\mathbf{z}\Vert_{2}^{2}
\end{equation}
Let $\mathbf{z} = \mathbf{u}_{t}^{(k)}$, $\mathbf{y} = \mathbf{u}_{t}^{(k)}- \eta \frac{\partial J}{ \partial \mathbf{u}_{t}^{(k)}}$:
\begin{equation}
    \begin{aligned}
        J(\mathbf{u}_{t}^{(k)}- \eta \frac{\partial J}{ \partial \mathbf{u}_{t}^{(k)}}) &= J(\mathbf{u}_{t}^{(k)}) -\eta\Vert\frac{\partial J}{ \partial \mathbf{u}_{t}^{(k)}}\Vert_{2}^{2} + \frac{1}{2}L\Vert\eta\frac{\partial J}{ \partial \mathbf{u}_{t}^{(k)}}\Vert_{2}^{2} \\
        &= J(\mathbf{u}_{t}^{(k)}) -(\eta-\frac{1}{2}L\eta^{2})\Vert\frac{\partial J}{ \partial \mathbf{u}_{t}^{(k)}}\Vert_{2}^{2}
    \end{aligned}
    \label{eqLip}
\end{equation}
Since $0 < \eta \leq 1/L$, $\eta-\frac{1}{2}L\eta^{2}>0$, we have:
\begin{equation}
    J(\mathbf{u}_{t}^{(k)}- \eta \frac{\partial J}{ \partial \mathbf{u}_{t}^{(k)}}) < J(\mathbf{u}_{t}^{(k)})
\end{equation}
Hence, the gradient descent will decreases. Now we prove it will converge to global optimal solution $\mathbf{u}_{t}^{*}$. Since $J$ is convex \textit{w.r.t} $\mathbf{u}_{t}^{(k)}$, we have:
\begin{equation}
     J(\mathbf{u}_{t}^{(k)}) \leq J(\mathbf{u}_{t}^{*}) + \frac{\partial J}{ \partial \mathbf{u}_{t}^{(k)}} (\mathbf{u}_{t}^{(k)} - \mathbf{u}_{t}^{*})
\end{equation}
Plug Eq. (\ref{eqLip}) in. And consider $0\leq L\leq 1/\eta$, $\eta-\frac{1}{2}L\eta^{2}\geq\frac{1}{2}\eta$:
\begin{equation}
\begin{aligned}
    &J(\mathbf{u}_{t}^{(k)}- \eta \frac{\partial J}{ \partial \mathbf{u}_{t}^{(k)}}) \leq  J(\mathbf{u}_{t}^{(k)}) - \frac{1}{2}\eta\Vert\frac{\partial J}{ \partial \mathbf{u}_{t}^{(k)}}\Vert_{2}^{2} \\
    &\leq J(\mathbf{u}_{t}^{*}) + \frac{\partial J}{ \partial \mathbf{u}_{t}^{(k)}} (\mathbf{u}_{t}^{(k)} - \mathbf{u}_{t}^{*}) - \frac{1}{2}\eta\Vert\frac{\partial J}{ \partial \mathbf{u}_{t}^{(k)}}\Vert_{2}^{2}
\end{aligned}
\label{eq:ineq}
\end{equation}
Reform Eq. (\ref{eq:ineq}):
\begin{equation}
    \begin{aligned}
        &J(\mathbf{u}_{t}^{(k)}- \eta \frac{\partial J}{ \partial \mathbf{u}_{t}^{(k)}}) - J(\mathbf{u}_{t}^{*}) \\
        &\leq \frac{1}{2\eta}(\Vert\mathbf{u}_{t}^{(k)} - \mathbf{u}_{t}^{*}\Vert_{2}^{2} - \Vert\mathbf{u}_{t}^{(k)} - \mathbf{u}_{t}^{*} - \eta\frac{\partial J}{ \partial \mathbf{u}_{t}^{(k)}}\Vert_{2}^{2})
    \end{aligned}
\end{equation}
Summation over iterations:
\begin{equation}
    \begin{aligned}
        &\sum_{i=0}^{k-1} (J(\mathbf{u}_{t}^{(i)}- \eta \frac{\partial J}{ \partial \mathbf{u}_{t}^{(i)}}) - J(\mathbf{u}_{t}^{*}) )\\
        &\leq \sum_{i=0}^{k-1} (\frac{1}{2\eta}(\Vert\mathbf{u}_{t}^{(i)} - \mathbf{u}_{t}^{*}\Vert_{2}^{2} - \Vert\mathbf{u}_{t}^{(i+1)} - \mathbf{u}_{t}^{*}\Vert_{2}^{2}))\\
        &\leq \frac{1}{2\eta}\Vert\mathbf{u}_{t}^{(0)} - \mathbf{u}_{t}^{*}\Vert_{2}^{2}
    \end{aligned}
\end{equation}
Since $J(\mathbf{u}_{t}^{(k)}- \eta \frac{\partial J}{ \partial \mathbf{u}_{t}^{(k)}}) = J(\mathbf{u}_{t}^{(k+1)})< J(\mathbf{u}_{t}^{(k)})$, we have:
\begin{equation}
    \begin{aligned}
        J(\mathbf{u}_{t}^{(k)}) - J(\mathbf{u}_{t}^{*}) &\leq \frac{1}{k}\sum_{i=0}^{k-1} (J(\mathbf{u}_{t}^{(i)}- \eta \frac{\partial J}{ \partial \mathbf{u}_{t}^{(i)}}) - J(\mathbf{u}_{t}^{*}) ) \\
        &\leq \frac{1}{2\eta k}\Vert\mathbf{u}_{t}^{(0)} - \mathbf{u}_{t}^{*}\Vert_{2}^{2}.
    \end{aligned}
    \label{eq:converge}
\end{equation}

Eq. (\ref{eq:converge}) implies the convergence to global optimal solution $\mathbf{u}_{t}^{*}$ with rate $\mathit{O}(1/\eta)$. 
\end{proof}

\textit{Theorem 2} shows the global convergence of gradient descent over objective $J$ when $J$ is convex \textit{w.r.t} control $\mathbf{u}_{t}^{(k)}$. Since \textit{Lemma 1} holds, it also proves the global convergence of PMP Gradient controller (Algorithm. \ref{al1}). Further studies imply that for some special non-convex function $J$ (\textit{e.g.} satisfying \textit{Kurdaya-Lojasiewicz} Inequalities \cite{khamaru2018convergence}) gradient descent (and hence PMP Gradient Controller) will also converge to global optimal, and it is proved that they are unlikely be trapped in saddle points \cite{lee2016gradient}. For other common non-convex cases, PMP Gradient Controller will only converge to local optimal.

Note that in our Neural-PMP, we use NN approximated system dynamics $f_{NN}$ in Hamiltonian when performing PMP-Gradient algorithm. Since neural networks with conventional activation are \textit{Lipschitz Continuous}, the error between neural-simulated and true partial derivatives $\Vert\frac{\partial f}{\partial \mathbf{u}_{t}^{(k)}}-\frac{\partial f_{NN}}{\partial \mathbf{u}_{t}^{(k)}}\Vert, \Vert\frac{\partial f}{\partial \mathbf{x}_{t}^{(k)}}-\frac{\partial f_{NN}}{\partial \mathbf{x}_{t}^{(k)}}\Vert$ are limited. Assume that:
\begin{equation}
    \begin{aligned}
        \Vert\frac{\partial f}{\partial \mathbf{u}_{t}^{(k)}}-\frac{\partial f_{NN}}{\partial \mathbf{u}_{t}^{(k)}}\Vert \leq \delta; \\
        \Vert\frac{\partial f}{\partial \mathbf{x}_{t}^{(k)}}-\frac{\partial f_{NN}}{\partial \mathbf{x}_{t}^{(k)}}\Vert \leq \mu;
    \end{aligned}
\end{equation}
Then based on Hamiltonian Eq. \eqref{eqHam}, PMP Conditions Eq. \eqref{eqCostateForm}, \eqref{eqHamiltonMinimize}, it is easy to find that the error of gradient $\Vert\frac{\partial H}{\partial \mathbf{u}_{t}^{(k)}}-\frac{\partial H_{NN}}{\partial \mathbf{u}_{t}^{(k)}}\Vert$ in step (7) of Algorithm \ref{al1} should satisfy:
\begin{equation}
    \Vert\frac{\partial H}{\partial \mathbf{u}_{t}^{(k)}}-\frac{\partial H_{NN}}{\partial \mathbf{u}_{t}^{(k)}}\Vert \propto \mu^{T-1-t}\delta.
\end{equation}

Hence, the error between optimal solution obtained by Algorithm \ref{al1} under true dynamics and NN simulated dynamics $\Vert\mathbf{u}_{t}^{*} - \mathbf{u}_{t(NN)}^{*}\Vert$ should satisfy:
\begin{equation}
   \Vert\mathbf{u}_{t}^{*} - \mathbf{u}_{t(NN)}^{*}\Vert \propto M\mu^{T-1-t}\delta;
   \label{ErrorAnalysis}
\end{equation}
Where $M$ is the max iterations. When $f_{NN}$ is accurate enough to so that $\mu \to 0; \delta \to 0$, then the solution obtained by our Neural-PMP will be close to the solution under true dynamics.

\subsection{Settings of Neural-PMP for Problems with State and Action Bounds}
For optimal control problems with action bounds $\underline{u}, \overline{u}$ ($\underline{u}$: lower bound, $\overline{u}$: upper bound), we check the updated control action in Algorithm \ref{al1} at each iteration $k$:
\begin{equation}
    \mathbf{u}_{t}^{(k+1)} \leftarrow \mathbf{u}_{t}^{(k+1)} - \eta  \frac{\partial H(\mathbf{x}_{t}^{(k)}, \mathbf{u}_{t}^{(k)}, \mathbf{\lambda}_{t+1}^{(k)})}{\partial \mathbf{u}_{t}^{(k)}}
\end{equation}
Let $\mathbf{u}_{t}^{(k+1)}[i]$ denote element $i$ of obtained $\mathbf{u}_{t}^{(k+1)}$. We set:
\begin{equation}
     \mathbf{u}_{t}^{(k+1)}[i] \leftarrow \left\{
\begin{aligned}
&\overline{u}-\sigma & \text{if} \quad \mathbf{u}_{t}^{(k+1)}[i] \geq \overline{u}; \\
&\underline{u}+ \sigma   & \text{if}  \quad \mathbf{u}_{t}^{(k+1)}[i] \leq \underline{u}; \\
& \text{unchanged}, & \text{else};
\end{aligned}
\right.
\end{equation}
for each elements $i$ of $\mathbf{u}_{t}^{(k+1)}$ to ensure it is within bounds. $\sigma$ is a small value. We set it to be $10^{-6}$ in our experiments.

For optimal control problems with state bounds $\underline{x}, \overline{x}$ ($\underline{x}$: lower bound, $\overline{x}$: upper bound), we add an extra quadratic penalty term $P_e(\mathbf{x}_{t})$ in the running cost $L(\mathbf{x}_{t}, \mathbf{u}_{t})$:
\begin{equation}
    P_e(\mathbf{x}_{t})=\left\{
\begin{aligned}
&\sum_{i=1}^{m} \beta (\mathbf{x}_{t}[i]-\underline{x})^2  \qquad &\mathbf{x}_{t}[i]<\underline{x}, \\
&0  \qquad &0 \leq x_t \leq \overline{x}, \\
&\sum_{i=1}^{m} \beta (\mathbf{x}_{t}[i]-\overline{x})^2  \qquad &\overline{x} < \mathbf{x}_{t}[i].\\
\end{aligned}
\right.
\end{equation}

In the above penalty formulation, $\mathbf{x}_{t}[i]$ denotes element $i$ of $\mathbf{x}_{t}$. $m$ is the dimension of $\mathbf{x}_{t}$ and $\beta$ is the penalty coefficient. The penalty term $P_e$ generates huge cost when any element of $\mathbf{x}_{t}$ exceeds state bounds, serving as soft constraints for state variable $\mathbf{x}_{t}$.

\subsection{Detailed Settings of Experiment Environments}
\label{Systems}

We note that in our simulations, we observe that for some specific scenarios with strong local optima like Swimmer, lack of random search at the initial stage of policy update limits proposed method's performance. Moreover, for problems with long horizon and high dimension, the co-state recursion in PMP-Gradient may impose high computational burden. Below, we detail the simulation settings for the tested dynamical systems.

\noindent \textbf{Environment 1: Linear Quadratic Regulator (LQR)} 

Linear Quadratic Regulator (LQR) is a classical problem in optimal control theory. It has linear system dynamics and quadratic cost. A standard finite-horizon LQR can be formulated as:
\begin{equation}
    \begin{aligned}
        \min_{\{\mathbf{u}_{0:T-1}\}} \; & \sum_{t=0}^{T-1} (\mathbf{x}_{t}^{'}\mathbf{Q}\mathbf{x}_{t}+\mathbf{u}_{t}^{'}\mathbf{R}\mathbf{u}_{t}) + \mathbf{x}_{T}^{'}\mathbf{Q}_T\mathbf{x}_{T} \\ 
        s.t. \quad &{x}_{t+1}  = \mathbf{A}\mathbf{x}_{t} + \mathbf{B}\mathbf{u}_{t}.  
    \end{aligned}
\end{equation}
where $\mathbf{Q}$, $\mathbf{R}$ and $\mathbf{Q}_T$ are \textit{positive semi-definite} matrices. In our experiment, we set $\mathbf{Q}$ and $\mathbf{A}$ to be identity matrices with dimension 5, $\mathbf{R}$ to be identity matrix with dimension 3, and
\begin{equation}
    \mathbf{Q}_T = \begin{bmatrix}
        5 & 0 & 0 & 0 & 0 \\
        0 & 4 & 0 & 0 & 0 \\
        0 & 0 & 2 & 0 & 0 \\
        0 & 0 & 0 & 1 & 0 \\
        0 & 0 & 0 & 0 & 3 \\
    \end{bmatrix},  \; \mathbf{B} = \begin{bmatrix}
        1 & 0 & 0 \\
        0 & 1 & 0\\
        0 & 0 & 1 \\
        1 & 1 & 0 \\
        0 & 1 & 1 \\
    \end{bmatrix}
\end{equation}

For fitting of linearized and neural networks (NN) dynamics model, we randomly collected samples in action space $[-10^5, 10^5]$ to establish training dataset $\mathbb{D}=\{\mathbf{x}_{t}^{(i)}, \mathbf{u}_{t}^{(i)}| \mathbf{x}_{t+1}^{(i)}  \}_{i=1,2,...,N}$.

We fit 2-layer NN dynamics with ReLU activation by minimizing Eq. \eqref{equ:lossSys}, while we fit linearized model $\mathbf{x}_{t+1} = \widetilde{A}\mathbf{x}_{t} + \widetilde{B}\mathbf{u}_{t}$ by minimizing the mean squared error (MSE) using the same number of training samples as we use for Neural-PMP training:
\begin{equation}
    \min_{\widetilde{A}, \widetilde{B}} \; \sum_{i=1}^{N} (\Vert  \widetilde{A}\mathbf{x}_{t}^{(i)} + \widetilde{B}\mathbf{u}_{t}^{(i)} - \mathbf{x}_{t+1}^{(i)} \Vert_{2})^2
\end{equation}

In LQR, as the true dynamics is linear, both linearzied and NN models perform well for system identification.

\noindent \textbf{Environment 2: Battery} 

The optimal control task of Battery system involves a 1-dimensional control variable $\mathbf{u}_t$ ($\underline{u}\leq\mathbf{u}_t\leq \overline{u}$), which represents the amount of electricity charged or discharged at time step $t$; and a 1-dimensional state $\mathbf{x}_t$, which represents the remaining battery electricity. The running cost consists of three components: (1) charging cost $p_t\mathbf{u}_t$ (where $p_t$ is the time-varying grid electricity selling price) (2) penalty for excessive (dis)charging amount $\mathbf{u}_t$: $\alpha \mathbf{u}^{2}_t$ (3) penalty term $P_e(\mathbf{x}_t)$ as soft constraint to prevent $\mathbf{x}_t$ from exceeding the battery capacity limit $\overline{x}$:

\begin{equation}
    P_e(x_t)=\left\{
\begin{aligned}
&\beta x^2_t  \qquad &x_t<0, \\
&0  \qquad &0 \leq x_t \leq \overline{x}, \\
&\beta (x_t-\overline{x})^2  \qquad &\overline{x} < x_t.\\
\end{aligned}
\right.
\end{equation}
Terminal cost $\gamma(\mathbf{x}_T-\mathbf{x}_f)^{2}$ penalizes deviations from target final state $\mathbf{x}_f$. 

In realistic battery systems, charging/discharging is not ideal and incurs losses, resulting in a charging efficiency which is usually nonlinear and dependent on the battery's state-of-charge. Typically, the efficiency is smaller when the charging/discharging speed is higher. We denote the state-of-charge as the control variable $\mathbf{u}_t$ and represent the charging efficiency function as $\zeta(\mathbf{u}_t)$. Thus, the system dynamics of a realistic battery can be formulated as follows:
\begin{equation}
\label{equ:battery_efficiency}
    \mathbf{x}_{t+1}= \zeta(\mathbf{u}_t)\mathbf{u}_t + \mathbf{x}_t;
\end{equation}

Usually, each realistic battery has its own unique charging efficiency $\zeta(\mathbf{u}_t)$. In the simulation, we choose a nonlinear function $\zeta(\mathbf{u}_t) = 0.5 + \frac{1}{1+e^{\mathbf{u}_t}}$ as the ground-truth efficiency, which resembles many real-world batteries' charging efficiency curves. We note that such battery's charging efficiency function is not known explicitly by the users or controller, thus it is appropriate to use a neural network to approximate such dynamics \eqref{equ:battery_efficiency}.

The full formulation of the optimal control problem is:
\begin{equation}
    \begin{aligned}
        \min_{\{\mathbf{u}_{0:T-1}\}} \; & \sum_{t=0}^{T-1} (p_t\mathbf{u}_t + \alpha \mathbf{u}^{2}_t + P_e(\mathbf{x}_t)) + \gamma(\mathbf{x}_T-\mathbf{x}_f)^{2}  \\ 
        s.t. \quad &{x}_{t+1} = \zeta(\mathbf{u}_t)\mathbf{u}_t + \mathbf{x}_t  
    \end{aligned}
\end{equation}
In Battery environment, we fit linearized models and neural networks (NN) to learn part of system dynamics: the charging efficiency function $\zeta(\mathbf{u}_{t})$. We randomly collected samples in action space $[-5, 5]$ to establish training dataset $\mathbb{D}=\{\mathbf{u}_{t}^{(i)}| \zeta(\mathbf{u}_{t}^{(i)})  \}_{i=1,2,...,N}$. We train linearized model $\zeta(\mathbf{u}_{t}) = \widetilde{A}\mathbf{u}_{t}$ with MSE:
\begin{equation}
    \min_{\widetilde{A}} \; \sum_{i=1}^{N} (\Vert  \widetilde{A}\mathbf{u}_{t}^{(i)}  - \zeta(\mathbf{u}_{t}^{(i)})  \Vert_{2})^2
\end{equation}

Similarly, we train 2-layer NN dynamics with ReLU activation by MSE loss:
\begin{equation}
    \mathcal{L}_{sysid}=\frac{1}{N}\sum_{i=1}^{N} (f_{NN}(\mathbf{u}_{t}^{(i)}) - \zeta(\mathbf{u}_{t}^{(i)})^{2}.
\end{equation}

Since the true efficiency function is highly non-linear, linearized model failed to approximate it accurately. But NN models still perform well.

\noindent \textbf{Environment 3: Pendulum} 

For optimal control of Single Pendulum System, control variable $\mathbf{u}_{t}$ denotes the applied torque, while the 2-dimension state $[q_t, dq_t]^{'}$ includes angular velocity $q_t$ and angular acceleration $dq_t$.

Our objective is to minimize the summation of squared errors between $q_t$ and target $q_f$, $dq_t$ and target $dq_f$, and the square value of control torque $\mathbf{u}_{t}$ within time horizon $T-1$:
\begin{equation}
    \begin{aligned}
        \min_{\{\mathbf{u}_{0:T-1}\}} \; & \sum_{t=0}^{T-1} (w_q(q_t-q_f)^2+w_{dq}(dq_t-dq_f)^2+w_u\mathbf{u}_t^2 )\\ 
        &+ w_q(q_T-q_f)^{2}+w_dq(dq_T-dq_f)^{2} \\
        s.t. \quad &{x}_{t+1} = f(\mathbf{x}_{t}, \mathbf{u}_{t})
    \end{aligned}
\end{equation}

In the above formulation, $w_q$, $w_{dq}$, $w_u$ are cost weights. The system dynamics $f(\mathbf{x}_{t}, \mathbf{u}_{t})$ is established following the \textit{Newton's Second Law of Rotation}:
\begin{equation}
    f(\mathbf{x}_{t}, \mathbf{u}_{t}) = \mathbf{x}_t + \Delta [dq_t, \frac{u_t-mglq_t-\sigma sin(q_t)}{I}]^{'};
\end{equation}
where $m$ and $l$ denotes mass and length of the pendulum; $g$ is the gravity constant; $I=\frac{1}{3}mgl^2$ is the pendulum's moment of inertia; $\Delta$ is the time step length; and $\sigma$ is the damping ratio.

\end{document}